# Distance Optimal Formation Control on Graphs with a Tight Convergence Time Guarantee

Jingjin Yu    Steven M. LaValle

*Abstract*— For the task of moving a set of indistinguishable agents on a connected graph with unit edge distance to an arbitrary set of goal vertices, free of collisions, we propose a fast distance optimal control algorithm that guides the agents into the desired formation. Moreover, we show that the algorithm also provides a tight convergence time guarantee (time optimality and distance optimality cannot be simultaneously satisfied). Our generic graph formulation allows the algorithm to be applied to scenarios such as grids with holes (modeling obstacles) in arbitrary dimensions. Simulations, available online[1], confirm our theoretical developments.

## I. INTRODUCTION

In this paper, we study the problem of controlling a group of indistinguishable agents with non-negligible sizes to take arbitrary desired formations. The agents, confined to an arbitrary connected graph, are capable of moving from one vertex to an adjacent vertex in one time step. The control policy must ensure that no collisions occur, which may happen when two agents attempt to move to the same vertex or move along the same edge. Counting each edge as having unit distance, we show that a (centralized) policy/schedule exists that moves the agents to the desired formation along paths having shortest total distance. The control policy also guarantees that a convergence time (the time when the formation is complete) of no more than $n+\ell-1$, in which $n$ is the number of agents, $\ell$ is the maximum (shortest) distance between any two initial and goal vertices. Moreover, the algorithm for computing the policy has a time complexity of $O(nV^2)$, with $V$ being the number of vertices of the graph. This paragraph also summarizes the main contributions of this paper.

The general problem of formation control, sometimes also being referred to as rendezvous or consensus due to differences in emphases, has remained a central research topic in control theory and robotics; see, e.g., [1], [2], [3], [6], [9], [11], [13], [14], [15], [17], [20], [21], [23], [24], [25], [30]. An early account of the rendezvous problem, as a special case of formation control, appeared in [1], in which algorithmic solutions are provided for agents with limited range sensing capabilities. *Stop-and-go* strategies extending the algorithm in [1] are proposed in [13] and [14], which cover various synchronous and asynchronous formulations. An $n$-dimensional rendezvous problem was approached via proximity graphs in [3]. For the consensus problem it is shown that averaging the behavior of close neighbors causes all agents to converge to the same behavior eventually [9]. We point out that, although this paper works with disjoint initial and goal vertex sets of $n$ distinct elements each, the presented results can be easily generalized to any number of goal vertices between 1 and $n$, thus covering additional problems such as multi-agent rendezvous.

For the problem of achieving and maintaining formations in which not all agents are collocated, graph theoretic approaches are quite popular, probably because agents and inter-agent constraints can be represented naturally with vertices and edges of graphs [5], [23], [31]. On research that appears most related to our problem, a discrete grid abstraction model for formation control was studied in [16]. To plan the paths, a three-step process was used in [16]: 1) Target assignment, 2) Path allocation, 3) Trajectory scheduling. Although it was shown that the process always terminates, no characterization of solution complexity was offered. In contrast, we provide efficient algorithms that solve a strictly more general class of problems with optimality assurance. Our particular problem formulation is also closely related to the multi-robot path planning problem, studied actively in robotics [4], [8], [10], [18], [28], [19], [22], [26], [27]. In particular, we recently proposed a network flow based method for attacking the multi-agent path planning problem [29] (to be consistent with [29], the problem we study here is phrased as a multi-agent path planning problem). This paper, focusing on distance optimality and convergence time of the formation control problem, does not use a network flow based method. Some preliminary versions of the theoretical developments appeared in [29] without proofs. These proofs are provided in this paper.

The rest of the paper is organized as follows. Section II defines the problem of *formation control on graphs* and illustrates the problem and main results with an example. In Section III, we characterize the properties of a distance optimal path set, without explicit consideration of collision. Section IV shows that the optimal path set can be scheduled, free of collisions, with tightly bounded convergence time. In Section V, we present efficient algorithms that schedule distance optimal paths and discuss computation as well as simulation results. We conclude in Section VI.

This work was supported in part by NSF grants 0904501 (IIS Robotics) and 1035345 (Cyberphysical Systems), DARPA SToMP grant HR0011-05-1-0008, and MURI/ONR grant N00014-09-1-1052. We greatly appreciate the invaluable suggestions from the anonymous reviewers that helped improve the quality of the final presentation.

Jingjin Yu is with the Department of Electrical and Computer Engineering, University of Illinois at Urbana-Champaign, Urbana, IL 61801 USA. E-mail: jyu18@uiuc.edu. Steven M. LaValle is with the Department of Computer Science, University of Illinois at Urbana-Champaign, Urbana, IL 61801 USA. E-mail: lavalle@uiuc.edu.

[1]http://msl.cs.uiuc.edu/~jyu18/pe/formation.html.

## II. MODELING FORMATION CONTROL ON GRAPHS

### A. Formation Control on Graphs with Collision Prevention

Let $G = (V, E)$ be a connected, undirected, simple graph, in which $V = \{v_i\}$ is its vertex set and $E = \{(v_i, v_j)\}$ is its edge set. Let $A = \{a_1, \ldots, a_n\}$ be a set of agents that move with unit speeds along the edges of $G$, with initial and goal vertices on $G$ given by the injective maps $x_I, x_G : A \to V$, respectively. The set $A$ is effectively an index set. For convenience, we let $n = |A|$ and use $V, E$ to denote the cardinalities of the sets $V, E$, respectively, since the meaning is clear from the context. Let $\sigma$ be a permutation that acts on the elements of $x_G$, $(\sigma \circ x_G)$ is a map that defines a possible goal vertex assignment (a target formation).

A *scheduled path* is a map $p_i : \mathbb{Z}^+ \to V$, in which $\mathbb{Z}^+ := \mathbb{N} \cup \{0\}$. Intuitively, the domain of the paths is discrete time steps. A scheduled path $p_i$ is *feasible* for a single agent $a_i$ if it satisfies the following properties: 1) $p_i(0) = x_I(a_i)$. 2) For each $i$, there exists a smallest $k_{\min} \in \mathbb{Z}^+$ such that $p_i(k_{\min}) = (\sigma \circ x_G)(a_i)$ for some fixed $\sigma$ (i.e., same $\sigma$ for all $1 \le i \le n$). That is, the end point of the path $p_i$ is some unique goal vertex. 3) For any $k \ge k_{\min}$, $p_i(k) \equiv (\sigma \circ x_G)(a_i)$. 4) For any $0 \le k < k_{\min}$, $(p_i(k), p_i(k+1)) \in E$ or $p_i(k) = p_i(k+1)$. We say that two paths $p_i, p_j$ are in *collision* if there exists $k \in \mathbb{Z}^+$ such that $p_i(k) = p_j(k)$ (meet) or $(p_i(k), p_i(k+1)) = (p_j(k+1), p_j(k))$ (head-on). If $p(k) = p(k+1)$, the agent stays at vertex $p(k)$ between the time steps $k$ and $k+1$.

**Problem 1 (Formation Control on Graphs)** *Given a 4-tuple $(G, A, x_I, x_G)$, find a set of paths $P = \{p_1, \ldots, p_n\}$ and a fixed $\sigma$ such that $p_i$'s are feasible paths for respective agents $a_i$'s for this $\sigma$ and no two paths $p_i, p_j$ are in collision.*

In this paper, we assume that we work with graphs on which the only possible collisions that may happen are "meet" or "head-on" collisions. This assumption is a mild one: For example, a 2D grid with unit edge distance is such a graph for agents with radii of no more than $\sqrt{2}/4$.

### B. A Motivating Example

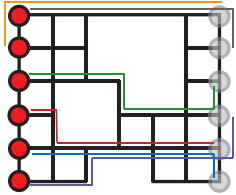

Fig. 1. A $6 \times 7$ grid with some vertices removed. The red discs represent the initial formation and the gray discs represent the goal formation.

To characterize what we solve in this paper, look at the example in Fig. 1. For the $6 \times 7$ grid with some vertices removed, we assign the top left corner coordinates $(0,0)$ and bottom right corner coordinates $(6,5)$. There are six agents with $x_I(A) = \{(0, i-1)\}, x_G(A) = \{(6, i-1)\}, 1 \le i \le 6$. That is, we want to move the agents from left to right. Given this problem, our algorithm first plans distance optimal paths without considering collisions (the multi-colored lines in Fig. 1). Then, the paths are ordered to produce the schedule in Table I. Each main entry of the table designates the coordinates an agent should be at the given time step. It is clear that a simple control policy can be easily generated from the schedule.

TABLE I

| Agent | Time Step | | | | | | | | |
|---|---|---|---|---|---|---|---|---|---|
| | 0 | 1 | 2 | 3 | 4 | 5 | 6 | 7 | 8 |
| 1 | 0,0 | 1,0 | 2,0 | 3,0 | 4,0 | 5,0 | 6,0 | 6,1 | 6,1 |
| 2 | 0,1 | 0,0 | 1,0 | 2,0 | 3,0 | 4,0 | 5,0 | 6,0 | 6,0 |
| 3 | 0,2 | 1,2 | 2,2 | 3,2 | 3,3 | 4,3 | 5,3 | 6,3 | 6,2 |
| 4 | 0,3 | 1,3 | 1,4 | 1,4 | 2,4 | 3,4 | 4,4 | 5,4 | 6,4 |
| 5 | 0,4 | 1,4 | 2,4 | 3,4 | 4,4 | 5,4 | 6,4 | 6,5 | 6,5 |
| 6 | 0,5 | 1,5 | 2,5 | 2,4 | 3,4 | 4,4 | 5,4 | 6,4 | 6,3 |

## III. SELECTING DISTANCE OPTIMAL PATHS

In this section, we pick a set of *unscheduled* paths $Q = \{q_1, \ldots, q_n\}$ (the colored paths in Fig. 1) that is *distance optimal* for the formation control task and characterize some of its properties. We use $Q$ to distinguish these paths from the *scheduled* paths, $P$. For convenience, $head(q_i)$, $tail(q_i)$, and $len(q_i)$ denote the start vertex, end vertex, and length of $q_i$, respectively. With a slight abuse of notation, $V(\cdot), E(\cdot)$ denote the vertex set and undirected edge set of the input parameter, which can be either a path, $q_i$, or a set of paths, such as $Q$. We define an *intersection* between two paths as a maximal consecutive sequence of vertices and edges common to the two paths.

Since we want to send agents from $x_I(A)$ to $x_G(A)$, we need a path set $Q$ such that $head(q_i) \in x_I(A), tail(q_i) \in x_G(A)$ for all $i$ and $|\{head(q_i)\}| = |\{tail(q_i)\}| = n$. It is clear that among all path sets satisfying the above property, there must be a set with the smallest total distance since there are only finitely many such path sets (there may be multiple such path sets with the same total distance). From this point onward, we use $Q$ to refer to an *arbitrary* unscheduled path set with shortest total distance, unless otherwise noted. This path set $Q$ has many interesting properties. Note that any path $q_i \in Q$ must be a shortest path between $head(q_i)$ and $tail(q_i)$. Once a $Q$ is selected, a $\sigma$ is implicitly determined.

**Lemma 2** *If we orient the edges of every path $q_i \in Q$ from $head(q_i)$ to $tail(q_i)$, then no two paths share a common edge of $E(Q)$ oriented in different directions.*

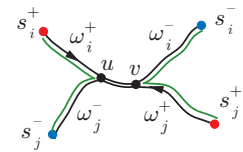

Fig. 2. Two opposite running paths, $q_i = s_i^+ \omega_i^+ uv \omega_i^- s_i^-$ and $q_j = s_j^+ \omega_j^+ vu \omega_j^- s_j^-$ (black paths), have total length at least 2 longer than that of $q_i' = s_i^+ \omega_i^+ \omega_j^- s_j^-$ and $q_j' = s_j^+ \omega_j^+ \omega_i^- s_i^-$ (green paths).

PROOF. Suppose the statement is false and assume that two oriented paths $q_i, q_j$ run in different directions on some common edge $(u, v)$. We may write the paths as $q_i = s_i^+ \omega_i^+ uv \omega_i^- s_i^-$ and $q_j = s_j^+ \omega_j^+ vu \omega_j^- s_j^-$, in which $\omega_i^+$ is the path of $q_i$ connecting $s_i^+$ to $u$ (see Fig. 2). $\omega_i^-, \omega_j^+, \omega_j^-$ are interpreted similarly. Then, the paths $q_i' = s_i^+ \omega_i^+ \omega_j^- s_j^-$ and $q_j' = s_j^+ \omega_j^+ \omega_i^- s_i^-$ have total length equaling $len(q_i) + len(q_j) - 2$, contradicting the minimality of $Q$. We conclude that no two oriented paths can have edges oriented in opposite directions. □

Above proof technique can be applied to show that $E(Q)$ can be oriented to form a *directed acyclic graph* (DAG).

**Theorem 3** *The path set $Q$ induces a DAG on $E(Q)$.*

PROOF. By Lemma 2, each edge of $E(Q)$ can be assigned a unique direction if we orient them from $head(q_i)$ to $tail(q_i)$ for $1 \le i \le n$. That is, the path set $Q$ induces a directed graph over $E(Q)$. Therefore, the claim of the theorem can only be false if there is a directed cycle in the induced graph. Since a single path from $Q$, being a shortest path, cannot form a directed cycle itself, at least two or more paths, say $q_1, \ldots, q_k$, are need to form a directed cycle. Without loss of generality, we assume these $k$ paths are all needed to form a cycle (i.e., $\{q_1, \ldots, q_k\} \setminus q_i$, $1 \le i \le k$, contains no directed cycle). That is, for each $1 \le i \le k$, the directed cycle, say $C$, has at least one edge that belongs only to $q_i$ (an illustration is given in Fig. 3). We show that we can update these paths, without changing the total distance of the path set, to obtain a path that intersects itself (containing a cycle). This means the total distance of the path set can be shortened by removing the cycle, a contradiction.

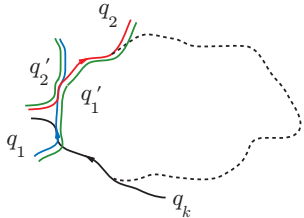

Fig. 3. A hypothetical (directed) cycle in a path set. We can switch the heads and tails of paths $q_1$ (blue) and $q_2$ (red) to get the (green) paths $q_1', q_2'$ without changing the total length. Now we can remove $q_2'$ and still have the same directed cycle. Performing the same procedure (essentially an inductive argument) eventually yields a path that contains a directed cycle.

We may write $q_1$ as $\omega_1 u \omega_2 v \omega_3$, in which $u\omega_2 v$ is the maximal segment of $q_1$ belonging to the cycle $C$; $\omega_1, \omega_3$ may be empty. Some other path intersecting $C$ must intersect $u\omega_2 v$ at $v$ (by the maximality of $u\omega_2 v$) and have a segment belonging to $C$ starting at vertex $v$; let $q_2$ be such a path. Since $q_2$ contributes some unique edges to $C$, there are some edges of $q_2$ in $C$ that follow $v$ but do not belong to $u\omega_2 v$. We can then write $q_2 = \omega_4 v \omega_5 w \omega_6$, in which $w$ is the last vertex of $q_2$ belonging to $C$. Note that $u\omega_2 v$ and $\omega_4 v$ may have edges that overlap. We can rearrange $q_1, q_2$ into $q_1' = \omega_1 u \omega_2 v \omega_5 w \omega_6$ and $q_2' = \omega_4 v \omega_3$. Clearly, $len(q_1) + len(q_2) = len(q_1') + len(q_2')$; the new path set is still minimal. We have shown that a path set $\{q_1, \ldots, q_k\}$ with a directed cycle can be rearranged to yield a path set such that $\{q_1', q_3, \ldots, q_k\}$ again contains the same directed cycle. Applying the same reasoning recursively yields a shortest path that contains the (same) directed cycle. □

If a vertex $v \in x_G(A)$ is on exactly one path $q \in Q$, $v$ is a *standalone* goal vertex. Theorem 3 implies the following.

**Corollary 4** *$Q$ has a standalone goal vertex.*

PROOF. At least one vertex $v \in x_G(A)$ must be a standalone goal vertex; otherwise, every goal must be on another path and the directed path containing the goals must close to form a directed cycle because the number of goals is finite, contradicting Theorem 3. □

## IV. SCHEDULING DISTANCE OPTIMAL PATHS

In this section we show that an arbitrary unscheduled path set $Q$ can be turned into a scheduled path set $P$ with a tight convergence time guarantee. As mentioned in the introduction, time optimality is measured by the time it takes the last agent to reach its goal (some call this measurement the *makespan*). Our scheduling algorithm is quite simple. In what follows, $D_Q$ is the DAG induced by $Q$ on $E(Q)$.

**Schedule 5 (Sequential Transfer Schedule)** *For each time step $t = i$, $0 \le i \le n-1$, over all standalone goal vertices (Corollary 4 guarantees at least one exists), pick an initial vertex that is closest to one of these standalone goal vertices on $D_Q$. Denoting this pair of initial and goal vertices as $s_{i+1}^+, s_{i+1}^-$, let the agent on $s_{i+1}^+$ move to $s_{i+1}^-$ following an arbitrary directed path on $D_Q$ (there may be more than one such path). The path followed by the agent is $q_i'$ and the time parameterized path is $p_i$. Remove $s_{i+1}^+$ from $x_I(A)$ and $s_{i+1}^-$ from $x_G(A)$ and repeat the process for $t = i+1$.*

With Schedule 5, distance optimality is not violated and no two $p_i, p_j \in P$ may collide, as shown in the next two lemmas. The statement of Lemma 6 may feel counter-intuitive due to its recursive nature; the proof and figures should make things more clear. Note that as a path set $Q$ is updated, the associated $\sigma$ is also updated implicitly.

**Lemma 6** *There exists an ordered path set $Q = \{q_1, \ldots, q_n\}$ such that for any $1 \le i \le n$, restricting to $Q_i := \{q_i, \ldots, q_n\}$, among all possible paths connecting an initial vertex (of $Q_i$) to a standalone goal vertex (of $Q_i$) using directed edges from $D_Q$, $q_i$ is a shortest such path.*

PROOF. We begin with a path set $Q = \{q_1, \ldots, q_n\}$ and construct a new path set $Q' = \{q_1', \ldots, q_n'\}$ satisfying the desired property, using edges from $E(Q)$. By Corollary 4, there are one or more standalone goal vertices. Among all possible paths connecting initial vertices and standalone goals using directed edges of $D_Q$, we pick one of the shortest. This is $q'_1$. Note it is likely that $q'_1 \notin Q$, in which case we may assume $head(q'_1) = head(q_i)$ and $tail(q'_1) = tail(q_j)$ for some $q_i, q_j \in Q$. There are two possibilities: Either $E(q'_1) \subset E(q_i) \cup E(q_j)$ or $q'_1$ contains edges from some other paths. For the first case (Fig. 4(a)), rearranging the paths as shown in green does not change total path length. For the second case, we may assume that $E(q'_1) \setminus (E(q_i) \cup E(q_j))$ belong to some other paths $q_k$ (applying similar reasoning used in the first case, we can always get such a $q_k$ via switching heads and tails of paths without changing the total path length). The switching shown in Fig. 4(b) gives us a $q'_1$ without changing total path length. After updating $Q$ (now contains $q'_1$ as an element), we apply the same procedure to $Q \setminus \{q'_1\}$ and so on; the end result is a path set satisfying the desired property. □

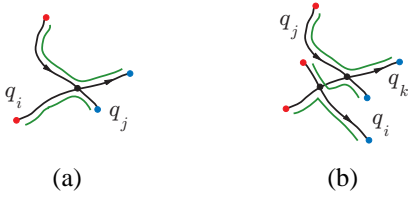

Fig. 4. Possible cases for rearranging paths without affecting total length. a) $q'_1$ is the lower green path. b) $q'_1$ is the middle green path.

**Lemma 7** *No two paths $p_i, p_j \in P$, produced by Schedule 5, will collide.*

PROOF. With a path set $Q$ produced by the construction used in Lemma 6, no path update is necessary; we are left to show that the scheduled paths will not collide. As stated in Section II, there are two types of collision for two scheduled paths $p_i, p_j$: Meet ($p_i(k) = p_j(k)$ for some $k$) and head-on $((p_i(k), p_i(k+1)) = (p_j(k+1), p_j(k))$ for some $k$). Lemma 2 rules out the possibility of having head-on collision. For the meet case, we prove via induction. For the base case, agent $a_1$ starts at $head(q_1)$ at $t = 0$. By construction, no other initial vertices can be closer to $tail(q_1)$ than $head(q_1)$. Since all other paths start later, they cannot get in the way of $q_1$'s schedule, which we denote $p_1$. Therefore, $p_1$ cannot collide with any other scheduled paths before it reaches its goal.

For the inductive case, assume that $\{q_1, \ldots, q_{k-1}\}$ can be scheduled to get $\{p_1, \ldots, p_{k-1}\}$ without collision. We need to show that $\{q_1, \ldots, q_k\}$ can be scheduled to get $\{p_1, \ldots, p_k\}$ without collision. Invoking the property that $tail(q_1)$ is a standalone goal vertex (that is, $p_1$ cannot collide with any other path on or after the time it reaches its goal, $tail(q_1)$), $q_1$ can be removed from the set $\{q_1, \ldots, q_k\}$ and induction hypothesis the applies to $\{q_2, \ldots, q_k\}$ to show that $\{p_2, \ldots, p_k\}$ contains no pairs that will collide. Adding $p_1$ back proves the inductive case. □

**Theorem 8** *Let $dist(u,v)$ denote the shortest distance between two vertices $u,v$. Schedule 5 provides a distance optimal solution to Problem 1 with convergence time no more than $n + \ell - 1$, in which*

$$\ell = \max_{u \in x_I(A), v \in x_G(A)} dist(u,v). \quad (1)$$

*Furthermore, the bound on the convergence time is tight.*

PROOF. Lemma 6 and 7 show that Schedule 5 correctly takes the agents to the desired formation. Since Schedule 5 handles one agent per time step, the last agent starts moving no later than $t = n - 1$ and finishes no later than $n + \ell - 1$.

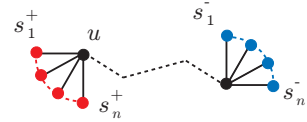

Fig. 5. An instance of Problem 1 for demonstrating the necessity claim of Theorem 8.

To see that the time bound $n + \ell - 1$ is necessary and therefore tight, look at an instance of Problem 1 shown in Fig. 5. The graph $G$ is two stars with centers connected by a single path; the red vertices form $x_I(A)$ and the blue ones $x_G(A)$. It is clear that all red vertices are of distance $\ell$ to all blue vertices. Given this graph $G$, only one agent can go from a red vertex to the adjacent black vertex $u$ in one time step. Thus, it takes at least $n$ time steps for the last agent at a red vertex to reach $u$. After that, it takes the last agent $\ell - 1$ steps to reach a blue vertex. Therefore, a total of $n + \ell - 1$ time steps is necessary. □

Moreover, in a sense, the convergence time given in Theorem 8 is the best we can hope for since distance optimality and time optimality cannot be simultaneously achieved.

**Proposition 9** *Distance optimality and time optimality for Problem 1 cannot be simultaneously satisfied.*

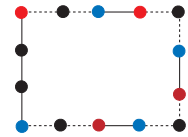

Fig. 6. A counter example showing that distance and time objectives cannot be optimized simultaneously.

PROOF. In Fig. 6, let the red (resp. blue) vertices be the initial (resp. goal) vertices. For distance optimality, the agents should take the solid paths with a total distance of $3 + 1 + 1 + 1 = 6$. These paths yield a value of 3 for the time objective since the longest path has a length of 3. If we optimize over time, then the dashed paths yield a value of 2 and they give a total distance of $2 + 2 + 2 + 2 = 8$. □

## V. COMPUTATIONAL COMPLEXITY AND RESULTS

### A. Pseudocode and Time complexity

Having proved that Schedule 5 takes $n$ agents to any formation within $n + \ell - 1$ time, we provide how such a schedule can be efficiently computed. The computation time will be given as worst case asymptotic bound in terms of the input parameters, $n, V$, and $E$. The scheduling routine, as described in Sections III and IV, is outlined in Algorithm 1.

**Algorithm 1** PLANANDSCHEDULEFORMATIONPATHS

**Input:** $G, A, x_I, x_G$ as input to Problem 1
**Output:** $P = \{p_1, \ldots, p_n\}$

1: **for** each $u_i \in x_I(A), v_j \in x_G(A)$ **do**
2:    obtain a shortest path $q_{ij}$ between $u_i, v_j$
3: **end for**
4: pick paths from $\{q_{ij}\}$ to form a path set $Q$
5: update $Q$ according to Lemma 6
6: **for** $t = 0$ to $n - 1$ **do**
7:    schedule $q_{t+1}$ to start at $t$ to get $p_{t+1}$
8: **end for**
9: **return** $P = \{p_1, \ldots, p_n\}$

In Algorithm 1, lines 1-3 can be realized with $n$ runs of breadth first search (BFS) on $G$, once for each $v \in x_I(A)$; this takes time $O(nE) \leq O(nV^2)$. A Hungarian algorithm [12] can then finish line 4 in $O(n^3)$ time. To compute an updated $Q$ (line 5), take $D_Q$ and invert the orientation of all its edges; denote the new graph $\overline{D_Q}$. We then create a new vertex $v_0$ and connect it to all standalone goals of $Q$ in $\overline{D_Q}$. Running BFS on $\overline{D_Q}$ from $v_0$ gives us an inverted $q'_1$ as constructed in the proof of Lemma 6. This is one iteration of Lemma 6, which takes time $O(E)$, resulting $O(nE) \leq O(nV^2)$ time for line 5. Lines 6-8 can be completed in $O(nE)$ time, which is bounded by $O(nV^2)$. Since $n \leq V$, the overall running time of Algorithm 1 is then $O(nV^2)$. This is faster than the network flow based algorithm from [29], which takes time $O(V^2 E \log V)$.

### B. Computational Results

We also evaluated the real world performance of the control strategy on commodity hardware[2]. In the evaluation, we used 2D grids as the underlying graph (for example, a 1600 vertex graph is a $40 \times 40$ grid) and randomly picked initial and goal vertices. The computational results are listed in Table II. The main entries are seconds that the algorithm takes to run for the given number of agents and vertices. The times are averages over 5 runs; the standard deviations are very small ($< 2\%$). "N/A" indicates that the number of agents are too many to put on the graph. Because all the subroutines used by Algorithm 1 are combinatorial [7] with small constants, even the Java implementation is fairly efficient on large graphs with many agents. For example, it takes 24 seconds to compute and schedule distance optimal paths for 1000 agents on a 10000 vertex graph.

TABLE II

| # Vertices | Number of Agents | | | | | | |
|---|---|---|---|---|---|---|---|
| | 10 | 25 | 50 | 100 | 250 | 500 | 1000 |
| 400 | 0.02 | 0.03 | 0.04 | 0.08 | N/A | N/A | N/A |
| 1600 | 0.03 | 0.03 | 0.08 | 0.14 | 0.52 | 1.98 | N/A |
| 10000 | 0.10 | 0.19 | 0.38 | 0.91 | 2.48 | 6.58 | 23.44 |
| 250000 | 4.64 | 8.31 | 16.01 | 30.8 | 78.21 | 188.0 | 391.2 |

### C. Heuristics and Simulations

Theorem 8 puts the worst case scheduling time bound at $n + \ell - 1$. However, if we adapt a simple heuristic, the total time to convergence can be greatly shortened. In many cases Schedule 5 can be compressed to yield a much shorter convergence time: Schedule a later path earlier when no conflict arises. We observed that when the agents and the goals are randomly scattered on a graph, more agents imply shorter convergence time (steps), as shown in Table III (for the simulation, a $21 \times 21$ grid was used. Initial and goal vertices were randomly picked; Fig. 7 captures one run with 75 agents. The data are averages over 10 runs). For example, the $n + \ell - 1$ bound translates to about 100 steps for 75 agents; our simulations show that on average only 10 steps are necessary. This is not surprising: When the graph is more crowded, the initial and goal vertices are generally closer.

TABLE III

| Number of Agents | 10 | 20 | 50 | 75 | 100 | 150 | 200 |
|---|---|---|---|---|---|---|---|
| Time Steps | 15.2 | 13.1 | 10.9 | 9.6 | 8.6 | 7.2 | 5.9 |

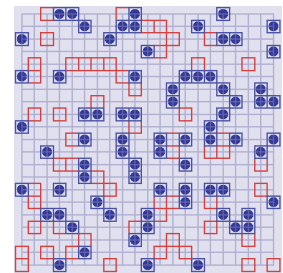

Fig. 7. A simulation capture. The red boxes are the initial vertices and the blue ones the goal vertices.

To confirm that Algorithm 1 is graph based and therefore insensitive to obstacles (as long as the obstacles are accounted for by the graph), we also examined grids with holes and observed no performance differences. The simulations mentioned in this subsection, as well as some additional interesting examples, are accessible on the Web (see abstract for the link).

## VI. CONCLUSION, FUTURE WORK, AND OPEN PROBLEMS

In this paper, we show that formation control on graphs, as defined in Problem 1, has distance optimal solutions that can be computed efficiently. Furthermore, the shortest paths can be scheduled to yield a control policy with a tight convergence time guarantee. The computation of the control policy can be carried out very efficiently.

---

[2]We implemented Algorithm 1 adhering to the Java 1.6 language standard under the Eclipse development environment. The computations were performed on a workstation with an Intel Core 2 Quad processor running at 3.0 GHz (only a single core was used). The JavaVM has a maximum memory of 3GB.

Two threads of future work are currently being explored, one of which is to make the algorithm decentralized, ideally requiring no global clock and only limited local communication. Another natural next step is to extend the results from graphs to continuous workspaces. There are many possible ways of doing this. For example, we may overlay a graph structure on an Euclidean space via discretization (see e.g., Fig. 8), which is not limited to grids. Given arbitrary initial and goal formations, we may design controllers by first aligning the initial and goal formations with vertices of the discrete graph. Algorithm 1 then applies. It remains to be characterized that how distance/time optimality might be affected and how differential constraints can be incorporated.

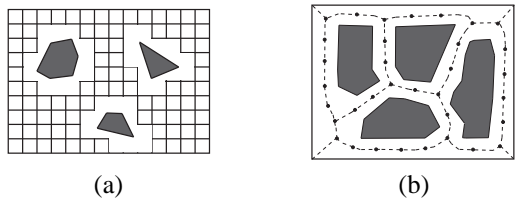

(a)          (b)

Fig. 8. Two types of discretizations.

Many open questions remain; we mention two here. In this paper, we only focused on indistinguishable agents. By restricting $\sigma$, different levels of distinguishability can be defined. For example, partitioning $\sigma$ into a few smaller ones with disjoint domains effectively grouping the agents into teams. In the extreme, $\sigma$ may be the identity map, assigning each agent a specific goal. It is interesting to see whether the structures enabled by shortest paths, present in Problem 1, generalizes to these problems. Shifting the emphasis to convergence time, we have demonstrated that it heavily depends on the graph structure, the number and the distribution of agents, and the distribution of goal vertices. Studying the interplay among these factors may lead to refined convergence time bound and better control policies.